\documentclass[a4paper]{article}

\usepackage{INTERSPEECH2020}
\usepackage{array,tabularx, url, caption}
\usepackage[super]{nth}
\usepackage[ruled,vlined]{algorithm2e}
\usepackage{silence}
\usepackage[]{microtype,multirow}

\title{FEARLESS STEPS Challenge (FS-2): Supervised Learning \\ with Massive Naturalistic Apollo Data}
\name{Aditya Joglekar, John H.L. Hansen, Meena Chandra Shekar, Abhijeet Sangwan}

\address{Center for Robust Speech Systems (CRSS),  Eric Jonsson School of Engineering, \\ The University of Texas at Dallas (UTD), Richardson, Texas, USA}
\email{\{aditya.joglekar, john.hansen, meena.chandrashekar, abhijeet.sangwan\}@utdallas.edu}

\begin{document}

\maketitle

\begin{abstract}
The Fearless Steps Initiative by UTDallas-CRSS led to the digitization, recovery, and diarization of 19,000 hours of original analog audio data, as well as the development of algorithms to extract meaningful information from this multi-channel naturalistic data resource. The 2020 FEARLESS STEPS (FS-2) Challenge is the second annual challenge held for the Speech and Language Technology community to motivate supervised learning algorithm development for multi-party and multi-stream naturalistic audio. In this paper, we present an overview of the challenge sub-tasks, data, performance metrics, and lessons learned from Phase-2 of the Fearless Steps Challenge (FS-2). We present advancements made in FS-2 through extensive community outreach and feedback.  We describe innovations in the challenge corpus development, and present revised baseline results. We finally discuss the challenge outcome and general trends in system development across both phases (Phase FS-1 Unsupervised, and Phase FS-2 Supervised) of the challenge, and its continuation into multi-channel challenge tasks for the upcoming Fearless Steps Challenge Phase-3.
\end{abstract}


\section{Introduction}
Recent decades have seen tremendous improvements to Speech and Language Technology (SLT) systems. This has only been possible due to thoroughly curated speech and language corpora that have been made publicly available~\cite{myPaper, fs1, ami, aspire, chimeCH}. The ability for systems to adapt to, and extract meaningful information from unlabeled data using limited ground-truth knowledge is a challenge in machine learning and AI~\cite{reverbCH, opensat, unsup}. Unfortunately, there is an unlimited amount of unstructured and unsupervised data compared to high quality human annotated data. To effectively address this reality, development of solutions will require consistent improvements to SLT systems. The initially digitized 19,000 hours from the NASA Apollo-11 and Apollo-13 missions~\cite{a11sol1, natArchiv} represent the largest naturalistic time synchronized multi-channel data. This corpus will be supplemented in continuing efforts with an additional 150,000 hours, enabling research on the largest publicly available corpus till date. Structuring this data through pipeline diarization transcripts, automatic speaker/sentiment tagging, etc., will enable preservation and archiving of historical data. These efforts will massively increase research opportunities, and be of significant benefit to the STEM community. 
As an initial step to motivate this stream-lined and collaborative effort from the SLT community, UTDallas-CRSS has been hosting a series of progressively complex tasks to promote advanced research on naturalistic “Big Data” corpora. This began with the “Inaugural FEARLESS STEPS Challenge: Massive Naturalistic Audio (FS-1)”. The first edition of this challenge encouraged the development of core unsupervised/semi-supervised speech and language systems for single-channel data with low resource availability, serving as the “First Step” towards extracting high-level information from such massive unlabeled corpora~\cite{fs1accept1, fs1accept2, fs1accept3, fs1accept4, fs1accept5}. As a natural progression following the successful inaugural FS-1 challenge, the FEARLESS STEPS Challenge Phase-2 (FS-2) focuses on the development of single-channel supervised learning strategies. 
FS-2 Challenge provides 80 hours of ground-truth data through training (Train) and development (Dev) sets, with an additional 20 hours of blind-set evaluation (Eval) data. Based on feedback from the Fearless Steps participants, additional tracks for streamlined speech recognition and speaker diarization have been included in the FS-2. To encourage diversified research interests, participants were also encouraged to utilize the FS-2 corpus to explore additional problems dealing with naturalistic data. The results for this challenge will be presented at the ISCA INTERSPEECH-2020 Special Session.

\vspace{-0.75em}
\section{Community Outreach \& Feedback}
The NASA Apollo Mission Control recordings are rich source of time-critical team based communications. Complex communication characteristics in this corpus can be explored through multiple avenues, and require vast resource utilization~\cite{myjasa, myconvjasa}. 
\vspace{-0.75em}
\begin{figure}[ht!]
  \centering
  \resizebox{4cm}{!}{
  \includegraphics[width=0.9\linewidth]{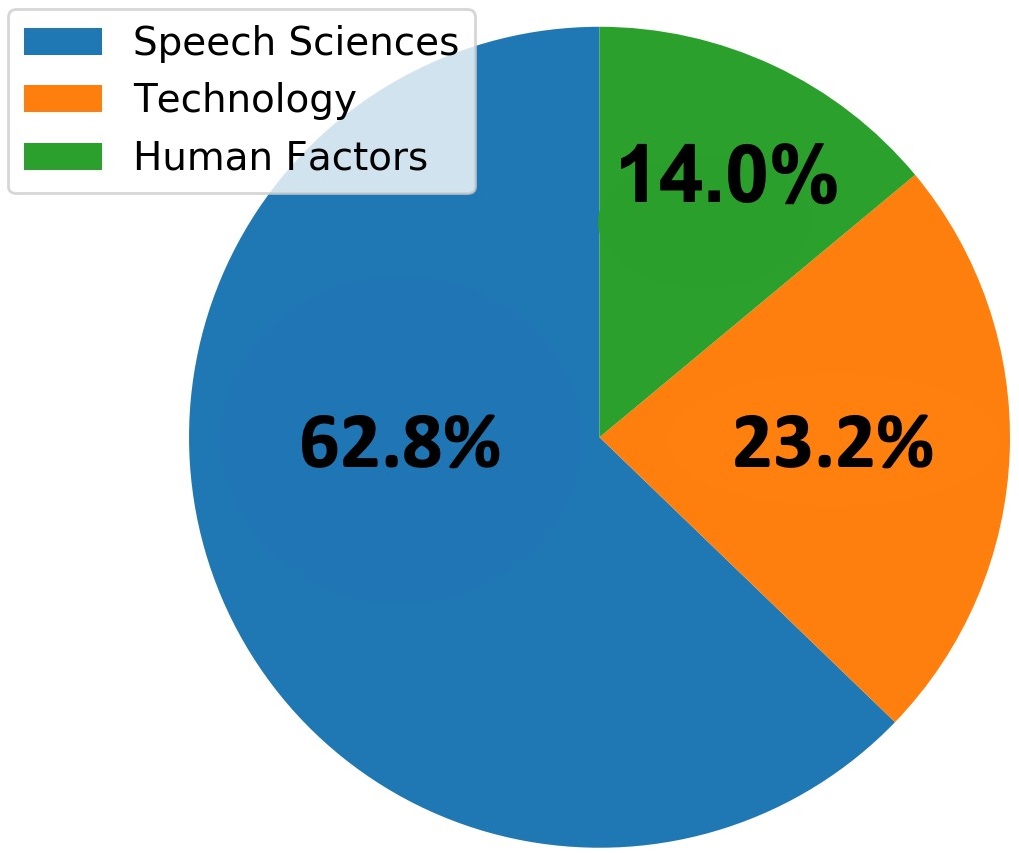}}
  \vspace{-2em}
\end{figure}
\begin{figure}[ht!]
  \centering
  \resizebox{8cm}{!}{
  \includegraphics[width=0.8\linewidth]{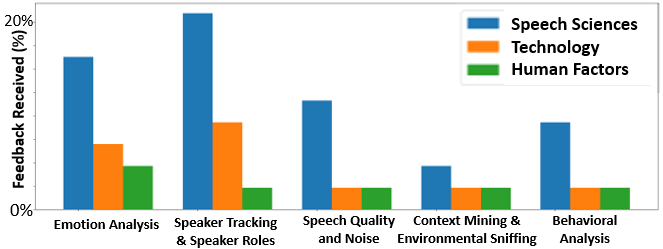}}
  \vspace{-1em}
  \caption{Analysis of community feedback. (top): Participant Breakdown  
  (bottom): Most requested areas of interest}
  \vspace{-0.75em}
  \label{fig:cf_bar_chart}
\end{figure}

To ensure optimum long-term benefits of exploring this corpus, feedback from researchers in multiple intersecting disciplines is crucial. An essential component of corpora development following the completion of FS-1 was a focus on community outreach and feedback. Multiple community engagement sessions were conducted with an aim in gathering essential future directions for the evolving FS Apollo corpus. The three communities directly benefiting from this corpus research and development include: (i) Speech Processing Technology (SpchTech), (ii) Communication Science and History (CommSciHist), and (iii) Education/STEM, Preservation/Archives, and Community-use (EducArch), who were consulted through workshop engagements. Community engagements are illustrated in Figure~\ref{fig:cf_bar_chart}.

\vspace{-0.5em}
\subsection{Fearless Steps Workshops}
\vspace{-0.25em}
User feedback was primarily collected through 6 workshops at STEM and archival events (including IS-19, JSALT-19, ASA-19, ASRU-19).  Online surveys of researchers downloading the FS corpus enabled access to feedback globally. A sample of the feedback from the above mentioned communities is illustrated in  Figure~\ref{fig:cf_bar_chart}.
The salient responses across all communities focused on availability of more labeled data for system development, linking unstructured audio data with relevant meta-data through robust semi-supervised SLT systems, convenient data access, and retrieval through pipeline diarization transcripts.

\vspace{-0.5em}
\subsection{Inaugural Fearless Steps (FS-1) Challenge}
\label{subsec:fs1}
\vspace{-0.25em}
Over 170,000 hours of synchronized audio data were collected by NASA during the Apollo missions. Digitizing this audio with synchronized SLT pipeline processing would enable streamlined information access and retrieval to all communities. Due to resource limitations on developing manual annotations, speech and language systems capable of extracting meaningful information using limited ground-truth resources are necessary. FS-1 was designed with this premise, providing 20 hours of development set ground-truth, and 20 hours of evaluation set for five tasks: Speech Activity Detection (SAD), Speaker Diarization (SD), Speaker Identification (SID), Speech Recognition (ASR), and Sentiment Detection. These 40 hours of data was selected from channels with comparatively lower levels of degradation. A lexicon and language model based on 4.2 billion NASA mission text content was also freely provided~\cite{srilm, lxkthesis}. Semi-supervised and unsupervised systems optimized for the Apollo data were used as baseline systems~\cite{combosad, kothapally2017speech, diarbaseline, fahimehSIDadapt, xia2019sid}. These systems have been used as benchmarks for evaluating the variability introduced in FS-2 by an additional 60 hours of audio from highly degraded channels. 

\vspace{-0.5em}
\begin{table}[htb]
\caption{Comparison of baseline results for FS-1 and FS-2 evaluation sets. Evaluation Metrics for FS-1 and FS-2: SAD: DCF (\%), SID: Top-5-Acc  (\%), SD: DER (\%), ASR: WER (\%), with Relative degradation in performance for same systems (\%)}
\vspace{-0.5em}
\centering
\resizebox{\linewidth}{!}{
\setlength{\tabcolsep}{8pt}
\begin{tabular}{|l|c|c|c|}
    \hline
    \multicolumn{4}{|c|}{\textbf{Fearless Steps System(s) Performance on Eval Set}}\\ \hline  
    \textbf{Task} & \textbf{FS-1 (\%)} &\textbf{FS-2 (\%)} & \textbf{Rel. Degradation (\%)}\\ \hline
    SAD & 11.70 & 13.60 & \textbf{16.20} \\
    SD & 68.23 & 88.27 & \textbf{29.37}\\
    SID & 47.00 & 41.70 & \textbf{11.27} \\
    ASR & 88.42 & 84.05 & - 4.90 \\
    \hline
\end{tabular}}
\vspace{-0.5em}
\label{tab:BaselineComparison}
\end{table}

With a goal to maintain competitiveness in FS-2, higher content of degraded audio was selected to form the Eval set in FS-2 to offset the advantage of Train set ground-truth availability. This is detailed in Section~\ref{subsec:select}. Table-\ref{tab:BaselineComparison} provides a comparison of baseline system performance for all tasks over the Eval sets of FS-1 and FS-2. Significant degradation in system performance in three out of four tasks is observed. The evaluation metrics used for tasks SAD, SD, SID, and ASR were detection cost function (DCF), diarization error rate (DER), top-5 accuracy (\%), and word error rate (WER) respectively~\cite{opensat, rttm, kaldi}.

Sentiment Detection task from FS-1 provided participants with rudimentary labels of `positive', `neutral', and `negative'. However, all communities expressed interest in descriptive labels for emotion and behavioral analysis, as seen in Figure~\ref{fig:cf_bar_chart}. Hence, sentiment detection task was removed from FS-2, and will be reintroduced in FS-3 as emotion detection task with 100 hours of improved labels.

\vspace{-0.5em}
\section{FS-2 Challenge Tasks}
The consensus from the community on requirement of increased transcribed data, and incremental task-targeted labeling prompted focused efforts on providing more variety in core-speech tasks. Hence, for FS-2, two separate challenge tracks were introduced for diarization and speech recognition. The speaker diarization track SD\_track2 focuses on developing robust speaker embedding and clustering algorithms, while SD\_track1 caters to the more challenging task of diarization from scratch. Equivalently, the Speech Recognition track ASR\_track2 focuses on transcribing diarized speech segments (each segment contains noisy speech from a single speaker), while ASR\_track1 incorporates the broader scope of transcribing noisy overlapped multi-speaker continuous streams. All challenge tasks for FS-2 are given in the following list: 

\begin{itemize}
\setlength\itemsep{0.75em}
\item \emph{TASK 1}: Speech Activity Detection \quad\quad\quad \textbf{\emph{(SAD)}}
\item \emph{TASK 2}: Speaker Identification \quad\quad\quad\quad\quad \textbf{\emph{(SID)}}
\item \emph{TASK 3}: Speaker Diarization 
\begin{list}{$\circ$}{\leftmargin=1em \itemindent=0em}
  \item (3.a.) \emph{Track 1}: using system SAD \quad\quad\space \textbf{\emph{(SD\_track1)}}
  \item (3.b.) \textit{Track 2}: using reference SAD \quad\space \textbf{\emph{(SD\_track2)}}
\end{list}
\item \emph{TASK 4}: Automatic Speech Recognition
\begin{list}{$\circ$}{\leftmargin=1em \itemindent=0em}
  \item (4.a.) Track 1: using system SAD \quad\quad \textbf{\textit{(ASR\_track1)}}
  \item (4.b.) Track 2: using diarized audio \quad\space \textbf{\emph{(ASR\_track2)}}
\end{list}
\end{itemize}
\vspace{0.25em}

The evaluation metrics for all tasks are consistent with the previous challenge, and described in Section~\ref{subsec:fs1}~\cite{kaldi, nistsre, dihard}. A scoring toolkit\footnote{\url{https://github.com/aditya-joglekar/FS02_Scoring_Toolkit}} was made publicly available for this challenge.

\vspace{-0.25em}
\section{Corpus Re-Deployment (FS-2)}

The five selected channels Flight Director (\textbf{FD}), Mission Operations Control Room (\textbf{MOCR}), Guidance Navigation and Control (\textbf{GNC}), Network Controller (\textbf{NTWK}), and Electrical Environmental and Consumables Manager (\textbf{EECOM}) from FS-1 were preserved with improved labeling for FS-2. The high degree of variability in speech and noise characteristics across these five channels has been explored previously~\cite{myPaper, fs1, lxkthesis, tocombo}. In FS-2, we introduce 60 hours of additional speech transcriptions and speaker labels from these channels to the existing 40 hours to provide sufficient data for supervised system training.

\begin{figure*}[htb!]
  \centering
  \includegraphics[width=0.9\textwidth]{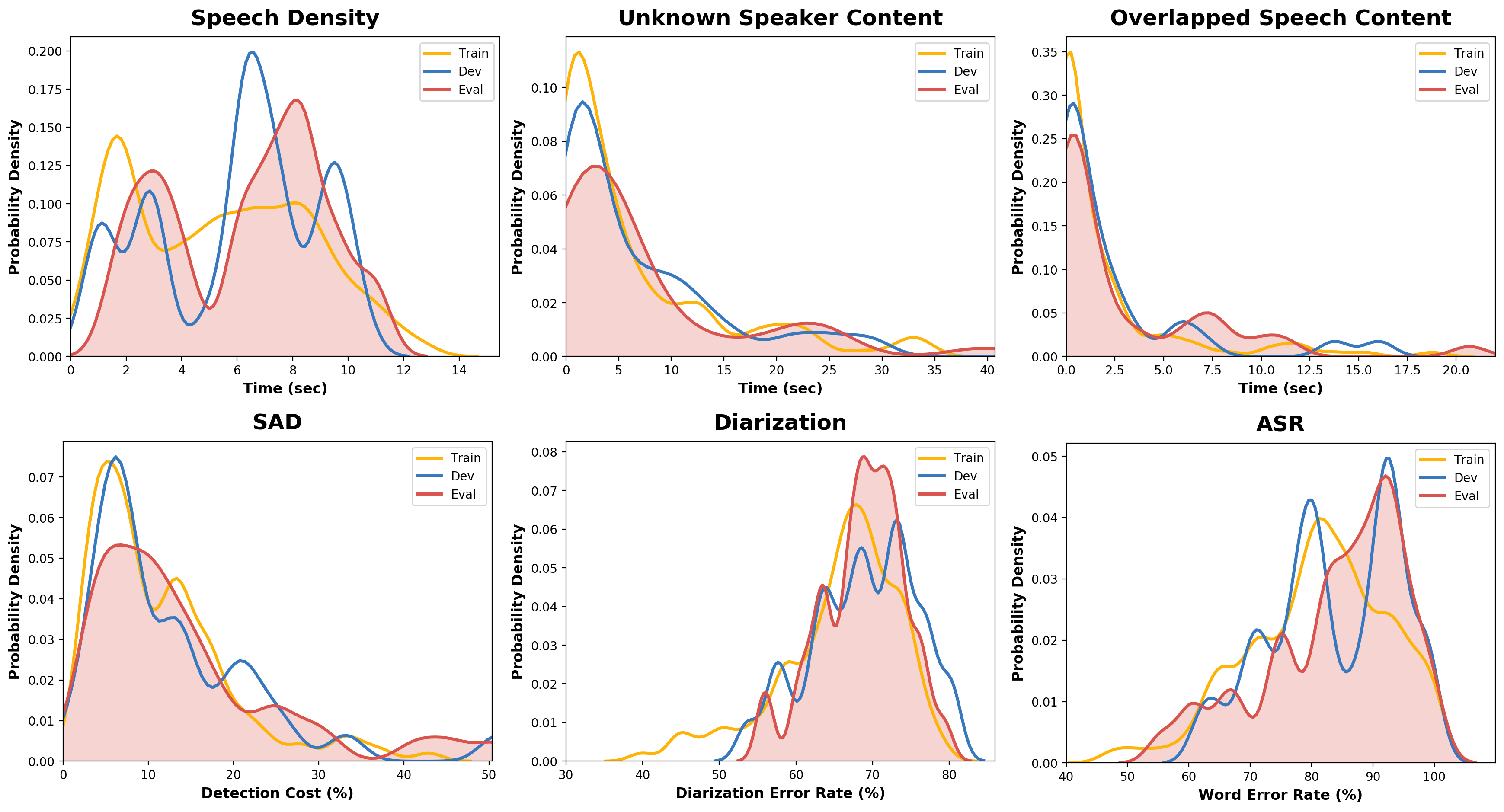}
  \vspace{-0.75em}
  \caption{Probability Distributions of decision parameters for Train, Dev, and Eval sets.}
  \label{fig:DistPlots}
\end{figure*}

\begin{table*}[htb!]
\captionsetup{justification=centering}
\caption{General statistics for the SID task. The mean, median, minimum, and maximum values \\for cumulative speaker durations, and individual speaker utterances are all expressed in seconds.}
\vspace{-0.5em}
\centering
\resizebox{0.9\textwidth}{!}{
\setlength{\tabcolsep}{8pt}
\begin{tabular}{|l|c|c|c|c|c|c|c|}
    \hline
    \multirow{2}{*}{\textbf{Data set}} & \multirow{2}{*}{\textbf{\# Spkrs}} & \multicolumn{3}{c|}{\underline{\textbf{Spkr. Duration (s)}}} & \multicolumn{3}{c|}{\underline{\textbf{Spkr. Utterances (s)}}} \\ 
    ~ & ~ & \multicolumn{1}{c}{\textbf{mean}} & \multicolumn{1}{c}{\textbf{median}} & \multicolumn{1}{c|}{\textbf{(min , max)}} & \multicolumn{1}{c}{\textbf{mean}} & \multicolumn{1}{c}{\textbf{(min , max)}} & \multicolumn{1}{c|}{\textbf{total}} \\ \hline
    Train  & 218 & 505.5 & 106.7 & (6.89 , 11254.36) & 4.03 & (1.84 , 16.95) & 27336\\ 
    Dev    & 218 & 118.1 & 24.2  & (3.13 , 2596.18) & 4.04  & (1.78 , 16.95) & 6373\\ 
    Eval   & 218 & 156.9 & 31.5  & (3.19 , 3460.41) & 4.04  & (1.8 , 16.22) & 8466\\ \hline
\end{tabular}}
\label{tab:SIDtab}
\vspace{-1em}
\end{table*}

\vspace{-0.5em}
\subsection{Data Set Selection}
\label{subsec:select}
The Dev, and Eval sets provided through FS-1 were developed using 70\% audio streams selected from clean channels, and 30\% selected from degraded channels. The Train, Dev, and Eval sets for FS-2 were categorized with scope to introduce multi-channel tasks in future challenges, while maintaining progressive difficultly in verification sets. The intention behind this data set design was to replicate naturalistic system development processes~\cite{chimeCH, reverbCH, bendre2020human}.
The FS-2 Challenge Corpus audio is divided into (i) audio streams, and (ii) audio segments. Audio streams reflect unaltered digitized audio from the Apollo missions. Audio segments are short duration speech sections diarized from the audio streams. Each segment contains a continuous speech utterance from a single speaker. Section~\ref{subsec:streams} describes the process of splitting 100 hours into Train, Dev, and Eval sets. Section~\ref{subsec:segments} provides more insight into development of segment based tasks SID and ASR\_track2.

\vspace{-0.5em}
\subsection{Audio Streams}
\label{subsec:streams}
Performance of SLT systems is dependent on factors like overlap content present in the data, amount of unintelligible speech, speech density variation, amount of data with unknown speakers, etc. In addition to this, the unsupervised baseline systems are useful in providing a measure of degradation in a given audio stream. We use the term 'decision parameters' to cumulatively describe the above measures. Using this methodology, it is possible to provide sets with progressive levels of difficulty across multi-channel audio streams in spite of inter-channel variations found in the Apollo data~\cite{myPaper}.
We perform this process by calculating all decision parameters for 100 hours of audio streams individually. These parameters are then normalized to generate degradation scores across the 100 hours. These scores are time-aligned across 5 channels and averaged to provide a single degradation score per 30-minute time chunk. These scores are finally categorized into three sets by progressive order of degradation. 5 channel segments with a cumulative highest degradation across all decision parameters are thus included in the Eval set, followed by Dev set. The streams with the least performance degradation are selected into the Train set. Trends observed from Figure~\ref{fig:DistPlots} explain that even when the overall degradation across multiple channels is large, due to the variances in channel characteristics, the distributions for Train, Dev, and Eval sets have similar means, but differing distributions. Such varying distributions across decision parameters can aid in assessing the robustness of systems and their ability to generalize to data with a high degree of cross-channel variability.

\begin{table}[htb!]
\caption{Duration Statistics of audio segments for ASR\_track2. The mean, min, and max values are expressed in seconds.}
\vspace{-0.75em}
\centering
\resizebox{\linewidth}{!}{
\setlength{\tabcolsep}{8pt}
\begin{tabular}{|l|c|c|c|c|}
    \hline
    \multirow{2}{*}{\textbf{Data set}} & \multirow{2}{*}{\textbf{Segments}} &  \multicolumn{3}{c|}{\underline{\textbf{Utterance Duration (s)}}}\\ 
    ~ & ~ & \multicolumn{1}{c}{\textbf{mean}} & \multicolumn{1}{c}{\textbf{min}} & \multicolumn{1}{c|}{\textbf{max}}\\ \hline
    Train  & 35,474 & 2.85 & 0.10  & 70.37 \\
    Dev    & 9,203 & 2.97 & 0.12 & 67.39  \\
    Eval   & 13,714 & 2.78 & 0.10  & 53.04 \\ \hline
\end{tabular}}
\label{tab:ASR_track2}
\vspace{-1em}
\end{table}

\vspace{-0.5em}
\subsection{Audio Segments}
\label{subsec:segments}
SID task in FS-1 challenge provided 183 speakers a minimum of 10 seconds of training data. FS-2 SID task extends this set by adding over 30,000 additional utterances for 218 speakers. With shorter utterance durations and larger variations in speaker durations as seen in Table-\ref{tab:SIDtab}, FS-2 provides a more challenging task over FS-1. This data also encapsulates the challenges faced in speaker tagging for Apollo corpora. While a few personnel had major speaking roles, most backroom staff in the mission control audio recordings had limited but integral speaking roles, making unbalanced and low resource speaker identification essential for a real-world scenario. Table-\ref{tab:ASR_track2} illustrates the general duration statistics of audio segments provided for the ASR\_track2 task. While this task has the advantage of having fully diarized segments, the single word utterance durations shorter than 0.2 secs pose a challenge to ASR systems.

\section{Baseline Systems}

The SAD, ASR, and speaker diarization baseline systems from the first challenge were retrained and optimized for usage in this challenge~\cite{fs1}. Both tracks for SD and ASR tasks were evaluated using the same system, with differing configurations. Baseline results for all tasks are provided in Table-\ref{tab:BaselineResults}. 

\vspace{-0.5em}
\begin{table}[htb!]
\caption{Baseline Results for Development and Evaluation Sets}
\vspace{-0.5em}
\centering
\resizebox{\linewidth}{!}{
\setlength{\tabcolsep}{12pt}
\begin{tabular}{|l|c|c|c|}
    \hline
    \multicolumn{4}{|c|}{\textbf{Fearless Steps Phase-02 Baseline Results}}\\ \hline 
    \textbf{Task} & \textbf{Metric} & \textbf{Dev (\%)} & \textbf{Eval (\%)}\\ \hline
     SAD & DCF & 12.50 & 13.60 \\
     SD\_track1 & DER & 79.72 & 88.27\\
     SD\_track2 & DER & 68.68 & 67.91\\
     SID & Top-5 Acc. & 75.20 & 72.46 \\
     ASR\_track1 & WER & 83.80 & 84.05 \\
     ASR\_track2 & WER & 80.50 & 82.23 \\
    \hline
\end{tabular}}
\vspace{-0.5em}
\label{tab:BaselineResults}
\end{table}

\vspace{-1em}
\subsection{Speaker Identification}

\begin{figure}[htb!]
  \centering
  \resizebox{8cm}{!}{
  \includegraphics[width=\linewidth]{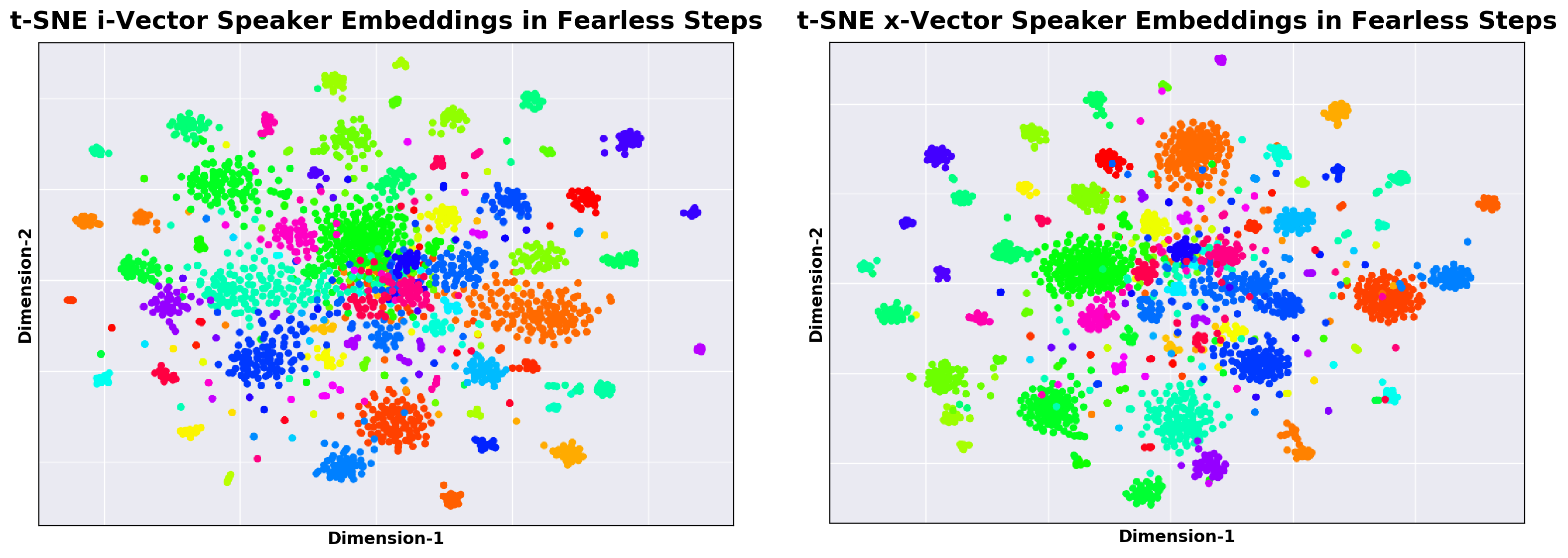}}
  \caption{Reduced dimensional i-Vector embedding \emph{(left)}, and x-Vector embedding \emph{(right)} t-SNE plots for 140 speakers~\cite{tsne}}
  \label{fig:tsne}
\end{figure}

The SID baseline system developed for FS-1 used i-Vectors for front-end processing~\cite{fahimehSIDadapt}. This system was more suited to the FS-1 SID data since it had at least 10 seconds of speech content per speaker. Due to the challenging nature of the current FS-2 SID data ($\leq$4 utterances per speaker on average), this system was rendered inadequate. Moreover, for speakers in the Apollo data, x-Vector and i-Vector embeddings have low separability, forming separate clusters for same speaker utterances from different channels. This is illustrated with a t-SNE plot of i-Vector and x-Vector embeddings for 140 speakers in Figure~\ref{fig:tsne}~\cite{ivector, xvector, tsne}.
\vspace{-0.5em}
\vspace{-0.5em}

\begin{figure}[htb!]
  \centering
  \resizebox{7cm}{!}{
  \includegraphics[width=0.75\linewidth]{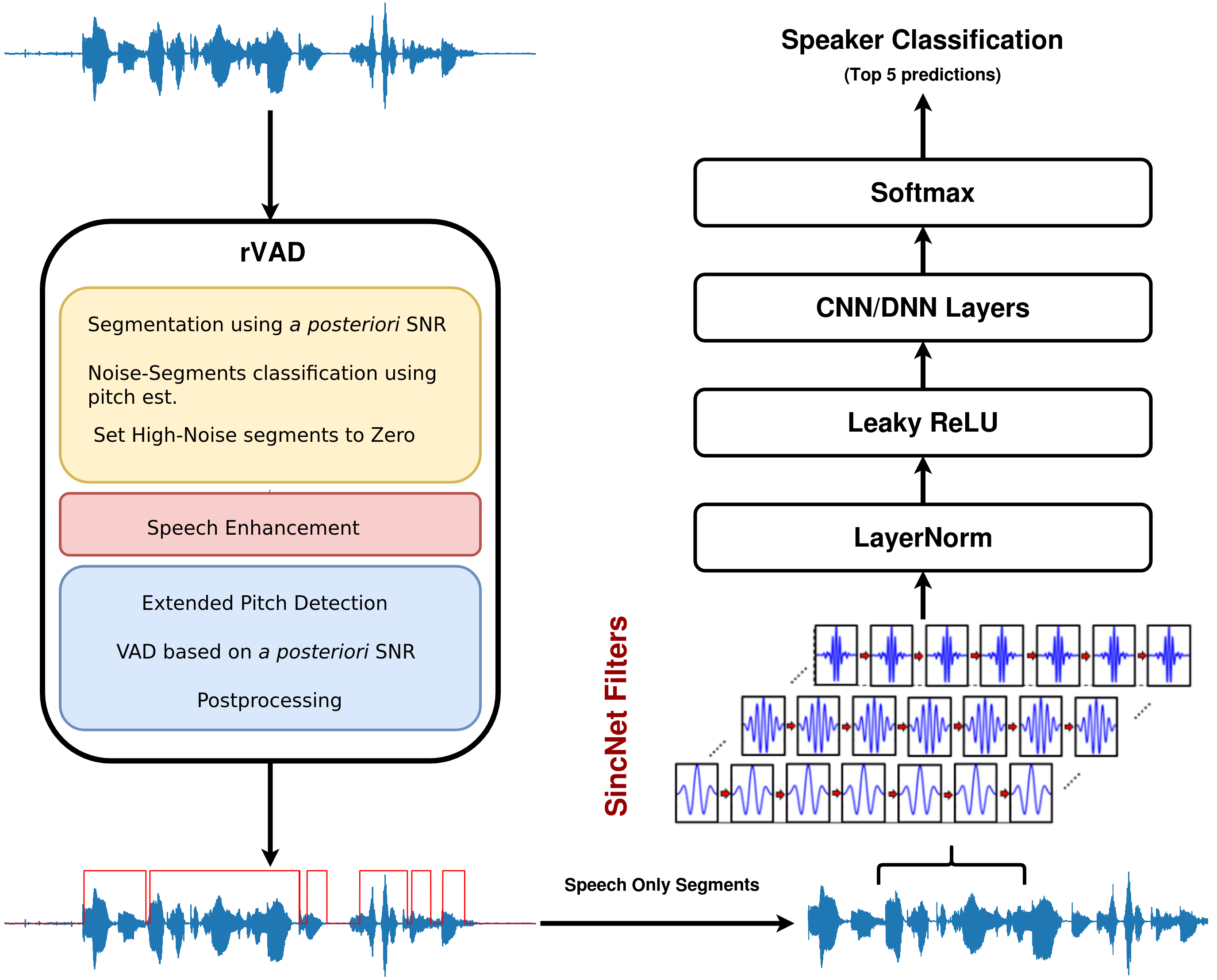}}
  \caption{rVad-SincNet based SID baseline system~\cite{SincNet, rVad}}
  \label{fig:sinc_net}
  \vspace{-1.5em}
\end{figure}

To provide an alternate baseline system more suited to the revised SID data, SincNet system was used~\cite{SincNet}. Input data was normalized and preprocessed to provide speech frames using the rVAD system (which ranked \nth{4} in the FS-1 SAD task)~\cite{rVad}. rVAD system threshold was optimized to provide strict speech boundaries. The SincNet was trained for 360 epochs. This system (shown in Figure~\ref{fig:sinc_net}) provided a Top-5 Accuracy of 72.46\%, which was a 30\% absolute improvement over the FS-1 SID baseline system.

\vspace{-0.75em}
\section{Discussion}
FS-2 Challenge concluded with 111 system submissions across all tasks. While this was similar to the 116 system submissions received for FS-1 challenge, participation for both tracks of SD and ASR tasks was noticeably higher. The systems developed for FS-2 also exhibited vast improvements in performance compared to the best systems developed for FS-1 challenge~\cite{fs1, fs1accept1, fs1accept2, fs1accept3, fs1accept5}, as seen in Table-\ref{tab:TopSystemComparison}. We observed relative improvements of 67\%, 57\%, and 62\% for SAD, Speaker Diarization from scratch, and Speech Recognition from audio streams tasks respectively. These top ranked systems from the community will be used to develop baselines for the next phase of the challenge, FS-3.

\vspace{-0.25em}
\begin{table}[htb!]
\caption{Comparison of the best systems developed for all FS-1 and FS-2 challenge tasks. Relative improvement of top-ranked system per task in FS-2 over FS-1 is illustrated.}
\vspace{-0.5em}
\centering
\resizebox{\linewidth}{!}{
\setlength{\tabcolsep}{12pt}
\begin{tabular}{|l|c|c|c|}
    \hline
    \multicolumn{4}{|c|}{\textbf{Comparison of Best System Submissions}}\\ \hline 
    \textbf{Task} & \textbf{FS-1 (\%)} & \textbf{FS-2 (\%)} & \textbf{Rel. Imp. (\%)}\\ \hline
     SAD & 3.31 & 1.07 & \textbf{\underline{67.67} \%} \\
     SID & 89.94 & 92.39 & \underline{2.72} \% \\
     SD\_track1  & 68.23 & 28.85 & \textbf{\underline{57.71} \%} \\
     SD\_track2  & \emph{N/A} & 26.55 & \emph{N/A} \\
     ASR\_track1 & 63.97 & 24.01 & \textbf{\underline{62.46} \%} \\
     ASR\_track2 & \emph{N/A} & 24.26 & \emph{N/A} \\
     \hline
\end{tabular}}
\label{tab:TopSystemComparison}
\vspace{-1em}
\end{table}

\vspace{-0.25em}
\section{Conclusions}

The FEARLESS STEPS Challenge Phases are aimed at developing robust speech and language systems for multi-party naturalistic audio. FS-2 enabled the development of new state-of-the-art supervised systems for core-speech tasks on Apollo data through its Challenge Corpus. Train, Dev, and Eval sets compatible for multi-channel challenges were also developed. Final Phase (FS-3) of the Fearless Steps initiative will include single and multi-channel core-speech tasks on the available 100 hours, and 20 hours of yet unrevealed Apollo-13 multi-channel audio (“Houston, we've had a problem”!). System advancements through FS-2 have also accelerated the development of conversational analysis and natural language understanding tasks for FS-3 like hot-spot detection, topic summarization, and emotion detection.

\vspace{-0.25em}
\section{Acknowledgements}

This project was supported in part by AFRL under contract FA8750-15-1-0205, NSF-CISE Project 1219130, and partially by the University of Texas at Dallas from the Distinguished University Chair in Telecommunications Engineering held by J.H. L. Hansen. We would also like to thank Tatiana Korelsky and the National Science Foundation (NSF) for their support on this scientific and historical project. A special thanks to Katelyn Foxworth (CRSS Transcription Team) for leading the ground-truth development efforts on the FS-2 Challenge Corpus. 

\newpage

\bibliographystyle{IEEEtran}

\bibliography{FS02_IS2020}

\end{document}